\documentstyle[12pt]{article}

\input epsf

\def\setb@se#1{\baselineskip=#1 \normalbaselineskip=#1}
\setb@se{14pt}
\newcommand{\be}{\begin{equation}}
\newcommand{\ee}{\end{equation}}
\newcommand{\g}{{\bf g}}
\newcommand{\x}{{\kappa}}

\newcommand{\A}{{\bf a}}
\newcommand{\C}{{s}}

\newcommand{\T}{\mbox{\bf T}}

\newcommand{\hhm}{\alpha}
\newcommand{\hhn}{\beta}
\newcommand{\hhp}{\gamma}
\newcommand{\hhq}{\delta}

\begin{document}
\bibliographystyle{plain}

\begin{titlepage}
\begin{flushright}

RU/97-5-B

ZU-TH/97-33

hep-th/9711181
\end{flushright}

\vspace{2 cm}

\begin{center}

{\huge Non-Abelian Solitons\\

\vspace{2 mm}

 in\\

\vspace{2 mm}

 N=4 Gauged Supergravity and \\

\vspace{4 mm}

Leading Order String Theory}

\vspace{20 mm}

{\bf Ali H. Chamseddine }\footnote{e-mail: chams@itp.phys.ethz.ch}

\vspace{1 mm}

{\it Institute of Theoretical Physics, ETH-H\"onggenberg,
CH--8093 Z\"urich, Switzerland}

\vspace{5 mm}

{\bf Mikhail S. Volkov}\footnote{e-mail: volkov@physik.unizh.ch}

\vspace{1 mm}

{\it  Institute of Theoretical Physics, University of Z\"urich,
Winterthurerstrasse 190, CH--8057 Z\"urich, Switzerland}

\vspace{5 mm}

\end{center}

\noindent
We study static, spherically symmetric, and purely magnetic
solutions of the N=4 gauged supergravity in
four dimensions. A systematic analysis of the supersymmetry conditions
reveals solutions which preserve 1/4 of the supersymmetries and are
characterized by a BPS-monopole-type
gauge field and a globally hyperbolic, everywhere regular geometry.
We show that the theory in which these solutions arise can be
obtained via compactification of ten-dimensional supergravity
on the group manifold. This result is then used to lift the
solutions to ten dimensions.

\vspace{30 mm}
\noindent
04.62.+v, 02.40.-k, 11.10.Ef, 11.25.-w, 11.30Pb
\end{titlepage}

\section{Introduction}

In the last few years there has been considerable interest in
supersymmetric solitons originating from
effective field theories of superstrings and heterotic strings
(see \cite{duff} for review).
These solutions play an important role in the study of the
non-perturbative sector of string theory and in understanding string
dualities. A characteristic feature of such solutions is that
supersymmetry is only partially broken, and associated
with each of the unbroken supersymmetries there is a Killing spinor
fulfilling a set of linear differential constraints.
The corresponding integrability conditions
can be formulated as a set of non-linear Bogomol'nyi
equations for the solitonic background, which can often
be solved analytically.

The analysis of the supersymmetry conditions has proven to be
an efficient way of studying the non-perturbative dynamics.
So far, however, the investigations
have mainly been restricted to the Abelian theory and
little is known about the structure of the non-Abelian sector, which
presumably is due to the complexity of the problem.
At the same time, the gauge group arising in the context
of string theory is fairly general.
It includes the  U(1) group as a subgroup,
but otherwise it is clear that the restriction
to the Abelian sector truncates most of the degrees of freedom.

In view of this it seems reasonable to focus on studying
supergravity solitons with non-Abelian gauge fields.
It turns out that all known solutions of this type can be
classified according to  two different methods
applied to obtain them.
The first of these methods is employed in
the heterotic five-brane construction \cite{strominger}.
Specifically,
the geometry of the four-dimensional space transverse to the brane
is supposed to be conformally flat.
This allows one to choose for the  Yang-Mills field
living in this space any know solution of the self-duality
equations in flat Euclidean space. Choosing all possible self-dual
configurations, one can obtain in this way a large variety of
different five-branes  \cite{HL},\cite{monopole}.
The ten dimensional solutions then further modify upon reducing to
four dimensions, displaying, nevertheless,
a number of common features
due to the common origin.
In this connection it is also worth mentioning the work in
Ref. \cite{GKLTT}, where
the equations of a supergravity model with non-Abelian vector fields
were directly attacked.
It was shown later \cite{GT1} that,
for one special value  of the dilaton coupling constant, 
the solutions obtained exactly correspond to the reduction
of the five-brane-type  solution described in \cite{HL}.

Another way to construct non-Abelian solutions is to
embed the gravitational connection into the gauge group;
see \cite{renata} and referencies therein. 
In this approach one starts 
from a solution of leading order string theory,
which is sometimes obtained by uplifting a
four-dimensional  Abelian solution. Its
spin-connections are then identified with the 
gauge field potential. As a result one obtains a solution 
of the theory with string corrections, which sometimes can 
be exact in all orders of string expansion. 

No other non-Abelian supergravity solitons are known than those
obtained by applying the described two methods.
All known solutions are thus either essentially Abelian,
or flat-space non-Abelian.
In this sense, they can be regarded as too special,
since none of them really
reflect the interplay between gravity and the non-Abelian gauge field.
At the same time, the famous  example of the
(non-supersymmetric) Bartnik-McKinnon particles \cite{BK}
shows that such an interplay can result in an
unusually rich variety of properties of the solutions.

Motivated by the arguments above, we study solitons in a
four-dimensional supergravity model
with non-Abelian Yang-Mills multiplets.
The model
we consider is the N=4 gauged SU(2)$\times$SU(2) supergravity \cite{FS},
which can be regarded as N=1, D=10 supergravity compactified on
the group manifold S$^{3}\times$S$^{3}$.
Note that all previously know non-Abelian supergravity solitons
have the Yang-Mills field
already in ten dimensions, and their compactification gives
the different matter content in D=4.
Our choice of the model therefore
ensures that we do not reproduce any known solutions.
Note also that the non-gauged version of the same model,
corresponding to the toroidal compactification of
ten-dimensional supergravity, has been
extensively studied in the past \cite{G}, \cite{LK}. 
The Abelian solutions in the gauged version of the model
have been studied in \cite{FG}.

In order to obtain the solutions we carry out the component
analysis of the supersymmetry constraints, which gives us a set
of the first integrals for the field equations. 
We investigate static, spherically
symmetric, purely magnetic field configurations
choosing for the gauge group either SU(2)$\times$SU(2) or
SU(2)$\times\left[{\rm U}(1)\right]^3$. It turns out that
in the first case there are no supersymmetric
solutions. The second choice, however, leads to the obtaining of
the variety of non-trivial solutions with 1/4 of  supersymmetries
preserved,  all of which can be described
analytically \cite{CV}. 
Among them we discover a one-parameter family
of globally regular solutions with quite unusual properties.
First, the solutions are characterized
by a regular-BPS-monopole-type gauge field with non-vanishing
magnetic charge.  This is very surprising,
since a Higgs field is not present in the problem,
in which case it would be reasonable to expect the regular
solutions to be neutral. Secondly, the geometry of the
solutions turns out to be globally hyperbolic.
This is also quite remarkable, because the standard gauge supergravity
ground states usually lack of global hyperbolicity.

Having obtained the solutions we lift them to ten dimensions.
For this we first show how to obtain the N=4
gauged supergravity  via compactification
of the N=1, D=10 supergravity on the group manifold,
which is either $S^3\times S^3$ or $S^3\times T^3$. 
It turns out that the corresponding procedure has not been described
in the literature.  Applying to the four dimensional solutions 
the procedure inverse to the dimensional reduction,
we thus obtain the solutions of the leading order equations of motion
of the string effective action in ten dimensions.

The rest of the paper is organized as follows.
In Section 2 we describe the action and supersymmetry
transformations of the N=4 gauged supergravity, derive the
field equations  and present their first integrals
following from the dilatational symmetry. Our procedure
to handle the supersymmetry constraints, that is, the
equations for the Killing spinors, is described in Section 3.
The supersymmetry conditions, which are the consistency conditions
for the supersymmetry constraints, are derived in Section 4
in the form of the first order Bogomol'nyi  equations for the
bosonic background. This section contains also the solutions
for the Killing spinors. Solutions of the Bogomol'nyi
equations are presented  in Section 5.
Section 6 describes the compactification of N=1, D=10
supergravity on the group manifold. The results obtained
in that section then used to lift the four-dimensional
solutions  to ten dimensions.
The lifted solutions and some of their properties are described
in Section 7. The last section contains concluding remarks. 

Our notation is as follows: Greek, Latin, and capital Latin
letters stand for the four-dimensional, internal six-dimensional,
and general ten-dimensional indices, respectively.
The early letters refer to the tangent space whereas the late
ones denote the base space indices. 
The six-dimensional space, whose indices are
a,b,c, $\ldots$
and $m,n,p, \ldots\, $, splits further into two three-dimensional
spaces. The three-dimensional indices are $a,b,c$, which stand
also for the group indices, and $i,j,k$.
The spacetime metric is denoted by $\g$, whereas $g$
stands for the gauge coupling constant(s).

\section{The model}
\setcounter{equation}{0}

The action of the N=4 gauged SU(2)$\times $SU(2) supergravity
includes a vierbein $e_{\mu }^{\hhm}$,
four Majorana spin-3/2 fields
$\psi_{\mu }\equiv \psi _{\mu }^{\rm{I}}$ $(\rm{I}=1,\ldots 4)$,
vector and pseudovector non-Abelian gauge fields
$A_{\mu }^{(1)\, a}$ and $A_{\mu }^{(2)\,a}$
with independent gauge coupling constants $g_{1}$ and $g_{2}$,
respectively,
four Majorana spin-1/2 fields $\chi \equiv \chi ^{\rm{I}}$,
the axion $\A$ and the dilaton $\phi$ \cite{FS}.
The bosonic part of the action reads
$$
S=\int \left( -\frac{1}{4}\,R
+\frac{1}{2}\,\partial _{\mu }\phi \,\partial^{\mu }\phi
+\frac{1}{2}\, e^{-4\phi}\partial _{\mu }\A \,\partial^{\mu }\A
-\frac{1}{4}\,e^{2\phi }\, \sum_{s=1}^2
{F}_{\mu \nu }^{(\C)\, a}F^{(\C)\, a\mu \nu }\right.
$$
\be                                       \label{1}
\left.
-\frac{1}{2}\,\A\, \sum_{s=1}^2
{F}_{\mu \nu }^{(\C)\, a}\ast\! F^{(\C)\, a\mu \nu }
+\frac{g^2}{8}\, e^{-2\phi }\right) \sqrt{-\bf{g}}\,d^{4}x.
\ee
Here $g^2=g_{1}^2+g_{2}^2$,
the gauge field tensor $F_{\mu \nu }^{(\C)\, a}=
\partial _{\mu }A_{\nu }^{(\C)\, a} -\partial _{\nu }A_{\mu}^{(\C)\, a}
+g_\C\, \varepsilon _{abc}\, A_{\mu }^{(\C)\, b}A_{\nu }^{(\C)\, c}$
(there is no summation over $\C=1,2$),
and $\ast\! F^{(\C)\, a}_{\mu\nu}$ is the dual tensor.
The dilaton potential can be viewed as an effective negative,
position-dependent cosmological term
$\Lambda(\phi)=-\frac{1}{8}\, g^2\, e^{-2\phi}$.
The ungauged version of the theory corresponds to the case
where $g_1=g_2=0$.

For a purely bosonic
configuration, the supersymmetry transformation laws are \cite{FS}
$$
\left.
\delta \bar{\chi}=
\frac{i}{\sqrt{2}}\,\bar{\epsilon}\,\right(
- \partial _{\mu }\phi +i\gamma_5\, e^{-2\phi}\, \partial_{\mu}\A\left)
\gamma^\mu
-\frac{1}{2}e^{\phi }\,\bar{\epsilon}\,
{\cal F}_{\mu \nu }\,\sigma ^{\mu \nu }+\frac{1}{4}\,e^{-\phi }\,
\bar{\epsilon}\, (g_1+i\gamma_5 \, g_2),\right.
$$
\begin{equation}                                  \label{2}
\delta \bar{\psi}_{\rho }=\bar{\epsilon}\left(
\overleftarrow{D}_{\rho }
-\frac{i}{2}\, e^{-2\phi}\, \partial_{\rho}\A\, \gamma_5\right)
-\frac{i}{2\sqrt{2}}\, e^{\phi }\,\bar{\epsilon}%
\,{\cal F}_{\mu\nu }\,\gamma_{\rho }\,
\sigma ^{\mu \nu }
+\frac{i}{4 \sqrt{2}}\, e^{-\phi }\,\bar{\epsilon}\,
(g_1+i\gamma_5 g_2)\, \gamma _{\rho },
\end{equation}
whereas the variations of the bosonic fields vanish. Here
$$
\bar{\epsilon}\overleftarrow{D}_{\rho}\equiv
\bar{\epsilon}\left( \overleftarrow{\partial}_{\rho }
-\frac{1}{2}\,\omega ^{\ \hhm\hhn}_{\rho}\,\sigma _{\hhm\hhn}
+\frac{1}{2}\, \sum_{s=1}^2  g_{\C}\, \T_{(\C)\, a}\, A_{\rho}^{(\C)\, a} \right),
$$
\be                                                \label{3}
{\cal F}_{\mu\nu}\equiv\T_{(1)\, a}\, F^{(1)\, a}_{\mu\nu}+
i\gamma_5\T_{(2)\, a}\,  F^{(2)\, a}_{\mu\nu}.
\ee
In these formulas,
$\epsilon \equiv \epsilon ^{\rm{I}}$ are four Majorana spinor
supersymmetry parameters, 
$\omega ^{ \ \hhm\hhn}_{\rho}$
 is the spin-connection, $\sigma _{\hhm\hhn}=\frac{1}{4}[
\gamma _{\hhm}\gamma_{\hhn}]$, 
and $\T_{(\C)\, a}\equiv\T_{(\C)\, a\, \rm IJ}$
are the SU(2)$\times$SU(2) gauge group generators,
whose explicit form will be given below.

Throughout this paper we shall specialize to the static,
purely magnetic fields. In this case the axion decouples and
one can consistently put $\A=0$. Choosing the metric in the form
\be                                          \label{4}
ds^2=e^{2V}dt^2-e^{-2V}h_{ik}\, dx^i dx^k,
\ee
the action becomes
$$
S=\int dt \int \left( -\frac{1}{4}\, ^{(3)}R
-\frac{1}{2}\,\partial _{i }V \,\partial^{i }V
-\frac{1}{2}\,\partial _{i }\phi \,\partial^{i }\phi
-\frac{1}{4}\,e^{2\phi+2V }\, \sum_{s=1}^2
{F}_{ik}^{(\C)\, a}F^{(\C)\, a ik}\right.
$$
\be                                         \label{5}
\left.
+\frac{g^2}{8}\,e^{-2\phi-2V }\right)
\sqrt{h}\,d^{3}x.
\ee
This admits a global symmetry
\be                                          \label{6}
V\rightarrow V+\epsilon,\ \ \ \
\phi\rightarrow \phi-\epsilon.
\ee
As a consequence, there exists a conserved Noether current
$\Theta^i=\sqrt{h}(\partial^i V-\partial^i\phi)$.
The corresponding conservation law is
\be                                  \label{7}
\tilde{\nabla}_i\tilde{\nabla}^i (V-\phi)=0,
\ee
where $\tilde{\nabla}_i$ is the covariant derivative
with respect to $h_{ik}$. As a result, the following condition
\be                                   \label{8}
V=\phi-\phi_0
\ee
with constant $\phi_0$ can be imposed.%

Let us now further specialize to the case of spherical symmetry.
For this we choose the spacetime metric and the gauge fields as
$$
ds^{2}=N\sigma ^{2}dt^{2}-\frac{dr^{2}}{N}-r^{2}(d\theta ^{2}+\sin
^{2}\theta \,d\varphi ^{2}),
$$
\be                                                \label{9}
\T_{(\C)\, a}A_{ \mu }^{(\C)\, a}dx^{\mu }=\left.\left.\left.\left.
\frac{1}{g_\C}\right(w_{\C}\,
\right\{-\T_{(\C)\, 2}\,d\theta 
+\T_{(\C)\, 1}\,\sin \theta \,d\varphi \right\}
+\T_{(\C)\, 3}\,\cos \theta \,d\varphi
\right) ,
\ee
where there is no summation over $\C$. We assume that the functions
$N$, $\sigma $, $w_\C$ and the dilaton $\phi $ depend only
the radial coordinate $r\in [0,\infty)$.
Substituting Eqs. (\ref{9}) into Eq. (\ref{1}) and omitting the surface
term, the action becomes
\be                                       \label{10}
\left. S=-4\pi\int dt\int_0^\infty dr\, \sigma\, \right\{
\frac{r}{2}\, (1-N)\, \frac{\sigma'}{\sigma} +\frac{r^2}{2}N\phi'^2
\left.
+\frac{1}{2}\, (NW+U)
-\frac{g^2}{8}\, r^2\, e^{-2\phi}
\right\},
\ee
where
\be                                         \label{10:1}
W\equiv W_1+W_2=2e^{2\phi}\sum_{\C=1}^{2}
\frac{ w_{\C}^{\prime 2}}{g_{\C}^2}
,\ \ \ \ \
U\equiv U_1+U_2=e^{2\phi}\sum_{\C=1}^{2}
\frac{(w_{\C}^2-1)^2}{g_{\C}^2\, r^2}.
\ee
This action admits a symmetry
\be                                       \label{11}
r\rightarrow \epsilon\, r,\ \ \
\sigma \rightarrow \frac{1}{\epsilon}\, \sigma,\ \ \
\phi\rightarrow\phi+\ln\epsilon,\ \ \
w_\C\rightarrow w_\C, \ \ \
N\rightarrow N,
\ee
which is the analog of that in Eq. (\ref{6}).
The corresponding  Noether current is
\be                                         \label{12}
\left. \Xi=\sum_{j}\frac{\partial L}{\partial u_j'}
\left(ru_j'-\frac{\partial u_j}{\partial\epsilon}\right)
\right|_{\epsilon=1}
-rL \equiv {\rm const.},
\ee
where $L=L(r,u_j,u_j')$ is the Lagrangian density corresponding to the
action (\ref{10}).

The field equations following from the action read
$$
(rN)^{\prime }+r^{2}N\phi ^{\prime \,\,2}+NW+U+r^2\Lambda(\phi)=1,
$$
$$
\sigma ^{\prime }/\sigma\, = r\phi ^{\prime \,\,2}+ W/r,
$$
$$
\left( \sigma Nr^{2}\phi ^{\prime }\right) ^{\prime
}=\sigma\, \{NW+U-r^2\Lambda(\phi)\},
$$
\be                                             \label{13}
\left( N\sigma e^{2\phi }\,
w_{\C}^{\prime }\right) ^{\prime }=\sigma e^{2\phi
}\,
w_{\C}(w_{\C}^{2}-1)/r^2.
\ee
Now, there are two first integrals for these equations
which provide the solution for the metric variables $N$
and $\sigma$. First, the condition (\ref{8}) ensures that
\be                                            \label{14}
\sigma^2 N=e^{2(\phi-\phi_0)}.
\ee
In addition, putting $\Xi=0$ in Eq. (\ref{12}) yields
\be                                             \label{15}
N=\frac{1-U
+g^2 r^2\, e^{-2\phi}/4}
{1+2r\phi'-r^2\phi'^2-W}.
\ee
 These two first integrals arise as a result of the
dilatational symmetry of the action. They provide the most
general solutions for the metric variables in the case where
the metric is regular at the origin. In addition,
as we shall see, these conditions are precisely what is
required by supersymmetry. One may wonder why the same
symmetry, being expressed in the two different
forms (\ref{6}) and (\ref{11}), leads to the two apparently
different expressions  (\ref{14}) and (\ref{15}).
It turns out that, although Eqs. (\ref{14}) and (\ref{15})
are indeed independent, they are equivalent up to an equation
of motion. Specifically, Eq. (\ref{15}) can be obtained
by inserting Eq.(\ref{14}) into the $G^r_r=2\, T^r_r$
Einstein equation.

Our  goal is to solve the remaining
equations in the system (\ref{13}). For this
we are turning now to the analysis of the supersymmetry
constraints, which will give us the additional
first integrals.


\section{Supersymmetry constraints}
\setcounter{equation}{0}

The field configuration (\ref{9}) is supersymmetric provided that
there are non-trivial
supersymmetry Killing spinors $\epsilon$ for which the variations of the
fermion fields defined by Eqs. (\ref{2}) vanish.
Putting in Eqs. (\ref{2}) $\delta\bar{\chi}=\delta\bar{\psi}_{\mu}=0$, 
we arrive at
the supersymmetry constraints given in the form of a system of
equations for the spinor supersymmetry parameter $\epsilon$ :
\be                                               \label{2.1}
2\sqrt{2}\, e^\phi\, \bar{\epsilon}\, \gamma^\mu\,\partial _{\mu }\phi
-2i\, e^{2\phi }\,\bar{\epsilon}\,{\cal F}_{\mu \nu }\,\sigma ^{\mu \nu }
+\bar{\epsilon}\, (i g_1-\gamma_5 \, g_2)=0,
\ee
\begin{equation}                                  \label{2.2}
4\sqrt{2}\, e^{\phi}\, \bar{\epsilon}\overleftarrow{D}_{\rho }
-2i\, e^{2\phi }\,\bar{\epsilon}%
\,{\cal F}_{\mu\nu }\,\gamma_{\rho }\,
\sigma ^{\mu \nu }
+\bar{\epsilon}\,
(i g_1-\gamma_5 g_2)\, \gamma _{\rho }=0.
\end{equation}
Here ${D}_{\rho}$ and ${\cal F}_{\mu\nu}$ are defined by
Eqs. (\ref{3}) and the background fields are specified by Eqs. (\ref{9}).
 This system consists of 80 linear equations for the 16
independent real components of $\epsilon$.
At most,
in the maximally supersymmetric case,
there could be 16
independent non-trivial solutions.
It is clear, however, that generically the system
has no non-trivial solutions at all. To find out under what
conditions the non-trivial solutions are possible,
our strategy is to analyse the equations in components.

First, we choose
the vierbein $e_{\hhm}^{\ \mu}$ to be a ``half-null'' complex tetrad:
\be                                             \label{2.3}
e_0=\frac{1}{\sigma\sqrt{N}}\, \frac{\partial}{\partial t},\ \ \
e_1=\sqrt{N}\, \frac{\partial}{\partial r},\ \ \
e_2=\frac{1}{\sqrt{2}r}\left(
\frac{\partial}{\partial\vartheta}
+\frac{i}{\sin\vartheta}\frac{\partial}{\partial\varphi}\right),\ \ \
e_3=e_{2}^{\ast}.
\ee
The non-zero components of the
tetrad metric $\eta_{\hhm\hhn}=(e_\hhm,e_\hhn)$ are
$\eta_{00}=-\eta_{11}=-\eta_{23}=1$.
The dual tetrad $e^\hhm$ determines the
spin-connection coefficients 
$\omega ^{\hhm\hhn}=
\omega ^{ \ \hhm\hhn}_{\rho}dx^\rho$
via the structure equation,
$de^\hhm+\omega^{\hhm}_{\ \ \hhn}\wedge e^\hhn=0$.

The gamma matrices 
$\gamma^\hhm\gamma^\hhn
+\gamma^\hhn\gamma^\hhm=2\eta^{\hhm\hhn}$
are chosen to be
$$
\gamma^0=
\left(\begin{array}{cc}
0 & 1 \\
1 & 0
\end{array}\right),\ \
\gamma^1=
\left(\begin{array}{cc}
0 & -\sigma^3 \\
\sigma^3 & 0
\end{array}\right), \ \
\gamma^2=\frac{1}{\sqrt{2}}
\left(\begin{array}{cc}
0 & -\sigma^{-} \\
\sigma^{-} & 0
\end{array}\right),
$$
\be                                      \label{2.4}
\gamma^3=\frac{1}{\sqrt{2}}
\left(\begin{array}{cc}
0 & -\sigma^{+} \\
\sigma^{+} & 0
\end{array}\right),  \ \
\gamma^5=
\left(\begin{array}{cc}
-1 & 0 \\
0 & 1
\end{array}\right),  \ \
C=
\left(\begin{array}{cc}
i\sigma^2 & 0 \\
0 & -i\sigma^2
\end{array}\right),
\ee
where 
$\gamma^5=\gamma_5=-\left(i/4!\right)\sqrt{-\eta}\, 
\varepsilon_{\hhm\hhn\hhp\hhq}\, 
\gamma^\hhm\gamma^\hhn
\gamma^\hhp\gamma^\hhq$ with $\varepsilon_{0123}=-1$
(note that $\sqrt{-\eta}=i$ since
$\det(\eta_{\hhm\hhn})=1$); 
and the charge conjugation matrix
$C\gamma^\hhm C^{-1}=-(\gamma^\hhm)^{\rm T}$.
The Pauli matrices are
$$
\sigma^1=
\left(\begin{array}{cc}
0 & 1 \\
1 & 0
\end{array}\right),\ \
\sigma^2=
\left(\begin{array}{cc}
0 & -i \\
i & 0
\end{array}\right), \ \
\sigma^3=
\left(\begin{array}{cc}
1 & 0 \\
0 & -1
\end{array}\right),
$$
and $\sigma^\pm=\sigma^1\pm i\sigma^2$.

The SU(2)$\times$SU(2) group generators $\T_{(\C)\, a}$,
which are subject to the conditions
$[\T_{(1)\, a},\T_{(2)\, b}]=0$ and
$\T_{(\C)\, a}\T_{(\C)\, b}=-\epsilon_{abc}\T_{(\C)\, c}-\delta_{ab}\, $,
are chosen to be
$$
\T_{(1)\, 1}=
\left(\begin{array}{cc}
0 & -\sigma^2 \\
\sigma^2 & 0
\end{array}\right),\ \
\T_{(1)\, 2}=
\left(\begin{array}{cc}
0 & -\sigma^1 \\
\sigma^1 & 0
\end{array}\right), \ \
\T_{(1)\, 3}=
\left(\begin{array}{cc}
-i\sigma^3 & 0 \\
0 & -i\sigma^3
\end{array}\right),
$$
\be                                       \label{2.5}
\T_{(2)\, 1}=
\left(\begin{array}{cc}
0 & i\sigma^3 \\
i\sigma^3 & 0
\end{array}\right),\ \
\T_{(2)\, 2}=
\left(\begin{array}{cc}
0 & -1 \\
1 & 0
\end{array}\right), \ \
\T_{(2)\, 3}=
\left(\begin{array}{cc}
-i\sigma^3 & 0 \\
0 & i\sigma^3
\end{array}\right).
\ee
Note that this representation of the group generators
differs from that in \cite{FS} by a unitary transformation.

The Majorana condition for $\epsilon$ requires that its Dirac
conjugate is equal to the
Majorana conjugate \cite{Nievenhuizen} :
\be                                             \label{2.6}
(\epsilon^{\rm I})^{\ast\rm T}\, \gamma^0
=\Omega^{\rm I}_{\ \rm J}\, (\epsilon^{\rm J})^{\rm T}C.
\ee
Here $\Omega^{\rm I}_{\ \rm J}$ is defined by the requirement that
the condition (\ref{2.6}) is invariant with respect to the
gauge transformations, which demands that
\be                                           \label{2.7}
\Omega\, \T_{(\C)\, a}+(\T_{(\C)\, a})^{\rm T}\, \Omega=0,
\ \ \ \Omega\, \Omega^\ast=1.
\ee
The solution of these equations,
in the representation  (\ref{2.5}),
is given by 
\be                                              \label{2.8}
\Omega=
\left(\begin{array}{cc}
\sigma^1 & 0  \\
0 & \sigma^1
\end{array}\right).
\ee
As a result, denoting the components of $\bar{\epsilon}^{\rm I}$ by
$\psi^{\rm I}_{q}$, the Majorana condition
can be expressed as a set of the following relations between
$\psi^{\rm I}_{q}$'s :
$$
\psi^{\rm 2}_1=-\left(\psi^{\rm 1}_4\right)^\ast,\ \
\psi^{\rm 2}_2=\left(\psi^{\rm 1}_3\right)^\ast,\ \
\psi^{\rm 2}_3=\left(\psi^{\rm 1}_2\right)^\ast,\ \
\psi^{\rm 2}_4=-\left(\psi^{\rm 1}_1\right)^\ast,\ \ \
$$
\be                                             \label{2.9}
\psi^{\rm 4}_1=-\left(\psi^{\rm 3}_4\right)^\ast,\ \
\psi^{\rm 4}_2=\left(\psi^{\rm 3}_3\right)^\ast,\ \
\psi^{\rm 4}_3=\left(\psi^{\rm 3}_2\right)^\ast,\ \
\psi^{\rm 4}_4=-\left(\psi^{\rm 3}_1\right)^\ast.
\ee

Now we can proceed to solving Eqs. (\ref{2.1}) and (\ref{2.2}).
First, we choose $\epsilon$ to be time independent. At this
stage one can obtain the first supersymmetry condition.
Specifically, let us multiply the $\rho=0$ equation in (\ref{2.2})
by $\gamma^0$ from the right and subtract the result from
Eqs. (\ref{2.1}). Using the fact that
the electric part of ${\cal F}_{\mu\nu}$ vanishes,
and also that $\gamma^0$
commutes with $\sigma^{ik}$,
the result is
\be                                          \label{2.10}
\bar{\epsilon}\gamma^\mu\partial_\mu\phi-
2\bar{\epsilon}\overleftarrow{D}_{0}\gamma^0=0.
\ee
Computing $\bar{\epsilon}\overleftarrow{D}_{0}
=-(1/2)\, \bar{\epsilon}\, 
\omega ^{\ \hhm\hhn}_{0}\,\sigma _{\hhm\hhn}$
this condition is equivalent to
\be                                         \label{2.11}
\bar{\epsilon}\gamma^1\left(\ln\left(\sigma^2 Ne^{-2\phi}\right)\right)'=0,
\ee
which finally requires that
\be                                          \label{2.11a}
\sigma^2 N=e^{2(\phi-\phi_0)},
\ee
thus reproducing Eq. (\ref{14}).
As a result, we can omit Eq.(\ref{2.1}) and concentrate on
the four gravitino supersymmetry constraints (\ref{2.2}).

Our procedure is straightforward: by inserting the above
definitions into Eqs. (\ref{2.2}) and projecting the equations
onto the tetrad,  we work out the result in components
(we do not present here the expressions explicitly
in view of their complexity).
The next step is to separate the angular variables,
and for this we take advantage of the special properties of the
spinor representation chosen.  Specifically,
it turns out that the spherical variables enter
the resulting equations only in such a way that they  form
certain differential operators. The structure of
these operators coincides with the one for the
raising and lowering operators
in the well-know recurrence relations
for the spin-weighted spherical harmonics $_{\x}Y_{jm}$ \cite{Gold} :
\be                                                 \label{2.14}
\left(\frac{\partial}{\partial\vartheta}
\mp\frac{i}{\sin\vartheta}\frac{\partial}{\partial\varphi}
\pm \x\cot\vartheta\right)\,  _{\x}Y_{jm}=\pm
\sqrt{(j\pm \x)(j\mp \x+1)}\,  _{\x\mp 1}Y_{jm}.
\ee
This suggests choosing
the spinor components $\psi^{\rm I}_q$
in the following form:
\be                                               \label{2.12}
\psi^{\rm I}_q\, (r,\vartheta,\varphi)
=R^{\rm I}_q(r)\, _{\x}Y_{jm}(\vartheta,\varphi).
\ee
The spin weights of the amplitudes, $\x=\x^{\rm I}_q$,
are determined by the
direct inspection of the equations:
$$
\x^{1}_1=\x^{1}_3=-\x^{2}_2=-\x^{2}_4=\frac{1-\nu_1-\nu_2}{2},\ \ \
\x^{1}_2=\x^{1}_4=-\x^{2}_1=-\x^{2}_3=-\frac{1+\nu_1+\nu_2}{2},
$$
\be                                             \label{2.13}
\x^{3}_1=\x^{3}_3=-\x^{4}_2=-\x^{4}_4=\frac{1-\nu_1+\nu_2}{2},\ \ \
\x^{3}_2=\x^{3}_4=-\x^{4}_1=-\x^{4}_3=-\frac{1+\nu_1-\nu_2}{2}.
\ee
Here $\nu_{\C}=1$ if $g_{\C}\neq 0$ and $\nu_{\C}=0$
otherwise.

The quantum number $j$, which is the same for all amplitudes,
has the meaning of the total
angular momentum including orbital angular momentum,
spin and isospin.  Its values are restricted by the
condition $j\geq |\x|$, since $_{\x}Y_{jm}$ vanishes otherwise.
We fix the value of $j$ by requiring that
\be                                                \label{2.15}
j=\min |\x ({\rm I},q)|,
\ee
where $\x({\rm I},q)$'s are given by Eq. (\ref{2.13}).
This can be regarded as a consistent truncation of the system,
since all amplitudes with $|\x({\rm I},q)|$ exceeding the
minimal value vanish.
The values of the azimuthal quantum number $m$
are restricted by the condition
 $-j\leq m\leq j$. Since $m$
does not enter the equations, its entire effect is to increase
the degeneracy of the solutions.

At this stage, the complete separation of the angular
variables is achieved in the equations.
The supersymmetry constraints reduce to a set
of algebraic and ordinary differential  equations
for the radial amplitudes $R^{\rm I}_{q}(r)$.
Note that the spin weights in Eq. (\ref{2.13}) and, correspondingly,
the structure of the resulting equations
essentially depend on whether
some of the coupling constants $g_\C$ vanish or not.
As a result, there arise three basically different cases to consider :\\
1) None of $g_\C$'s vanish, which corresponds to the full SU(2)$\times$SU(2)
gauge symmetry.\\
2) Either $g_1$ or $g_2$ vanishes -- the gauge symmetry is truncated to
SU(2)$\times\left[{\rm U(1)}\right]^3$.\\
3) $g_1=g_2=0$ -- the gauge group is $\left[{\rm U(1)}\right]^6$.\\
It turns out  that in the first case there are no solutions to the supersymmetry
constraints (apart from the trivial one).  If both coupling constants vanish, 
the non-trivial Killing spinors exist and 
the underlying supersymmetric backgrounds are
the well-known
Abelian dilaton black holes \cite{G}, \cite{LK}. Our main thrust
will be on the second case, where the gauge symmetry is truncated to 
SU(2)$\times\left[{\rm U(1)}\right]^3$.


\section{The supersymmetry consistency conditions.}
\setcounter{equation}{0}

If one of the coupling constants is zero, we assume that the
corresponding Abelian gauge field vanishes too.
At the same time, the other coupling constant can
be set to unity via the appropriate
rescaling of the fields in the action.
As a result, one has either $g_1=1$, $g_2=0$
or $g_1=0$, $g_2=1$. It turns out that
in both of these cases there is the same number
of non-trivial solutions of the supersymmetry constraints. 
The corresponding  consistency conditions
are identical up to the replacement
$w_1\leftrightarrow w_2$. We shall therefore consider explicitly 
only the case where $g_1=0$, $g_2=1$,
since the equations contain then only real coefficients.

Putting $A^{(1)\, a}_{\mu}=0$, the field equations
are obtained from Eqs. (\ref{13})--(\ref{15}) by
omitting the terms $W_1$ and $U_1$ in Eq.(\ref{10:1}).
The gauge field $A^{(2)\, a}_{\mu}$ is given by Eq. (\ref{9}),
where $w_2$ will be denoted by $w$.
 Eqs.(\ref{2.13}) imply that $\min |\x({\rm I},q)|=0$,
and so we put in (\ref{2.12}) $j=0$. Note that this can be regarded as a
manifestation of the spin-isospin coupling:
since both spin and isospin are half-integer,
the total angular momentum is integer and hence its
lowest value is zero. 
For $j=0$ all spin-weighted harmonics with $\x>0$ vanish,
while $_{0}Y_{00}=$const.
As a result, the non-vanishing spinor components are
\be                                             \label{3.1}
\bar{\epsilon}^{\, 1}=\left(R^{1}_1(r),0,R^{1}_3(r),0\right),\ \ \
\bar{\epsilon}^{\, 3}=\left(0,R^{3}_2(r),0,R^{3}_4(r)\right),\ \ \
\ee
and
\be                                             \label{3.1a}
\bar{\epsilon}^{\, 2}=\left(0,R^{2}_2(r),0,R^{2}_4(r)\right),\ \ \
\bar{\epsilon}^{\, 4}=\left(R^{4}_1(r),0,R^{4}_3(r),0\right).
\ee
Among these components
those in Eq. (\ref{3.1}) can be chosen to be independent,
whereas
\be                                             \label{3.1b}
R^{\rm 2}_2=\left(R^{\rm 1}_3\right)^\ast,\ \ \
R^{\rm 2}_4=-\left(R^{\rm 1}_1\right)^\ast,\ \ \
R^{\rm 4}_1=-\left(R^{\rm 3}_4\right)^\ast,\ \ \
R^{\rm 4}_3=\left(R^{\rm 3}_2\right)^\ast,
\ee
in view of the Majorana conjugation (\ref{2.9}).
The equations for $R^{2}_q$ and $R^{4}_q$ also  can be
obtained from those for $R^{1}_q$ and $R^{3}_q$
by applying the conjugation rule (\ref{3.1b}).
We shall therefore concentrate only on the independent
variables $R^{1}_q$ and $R^{3}_q$.

Making the linear combinations
\be                                        \label{3.2}
\Psi^1=R^{1}_1+R^{1}_3,\ \ \
\Psi^2=R^{3}_2+R^{3}_4,\ \ \
\Psi^3=R^{1}_1-R^{1}_3,\ \ \
\Psi^4=R^{3}_2-R^{3}_4,\ \ \
\ee
the supersymmetry constraints can be represented as follows:
The temporal component ($\rho=0$) of Eqs. (\ref{2.2}) 
gives the relations
$$
A^{+}\Psi^1+C\Psi^2=0,\ \ \
C\Psi^1-A^{-}\Psi^2=0,\ \ \
$$
\be                                       \label{3.3}
A^{-}\Psi^3-C\Psi^4=0,\ \ \
C\Psi^3+A^{+}\Psi^4=0,\ \ \
\ee
whereas the angular components of the equations 
($\rho=\vartheta, \varphi$) together require that
$$
b^{-}\Psi^1-w\beta\, \Psi^2=0,\ \ \
-w\beta\, \Psi^1+b^{+}\Psi^2=0,\ \ \
$$
\be                                       \label{3.4}
b^{+}\Psi^3-w\beta\, \Psi^4=0,\ \ \
-w\beta\, \Psi^3+b^{-}\Psi^4=0.\ \ \
\ee
Finally, the radial component yields
$$
\gamma\,(\Psi^{1})^{\prime}+(B+1)\Psi^{1}-C\Psi^{2}=0,\ \ \
\gamma\, (\Psi^{2})^{\prime}-(B+1)\Psi^{2}+C\Psi^{1}=0,\ \ \
$$
\be                                           \label{3.5}
\gamma\,(\Psi^{3})^{\prime}-(B+1)\Psi^{3}+C\Psi^{4}=0,\ \ \
\gamma\, (\Psi^{4})^{\prime}+(B+1)\Psi^{4}-C\Psi^{3}=0.\ \ \
\ee
The coefficients in these equations are given by
$$
B=\frac{2}{r^2}\, e^{2\phi}(w^2-1),\ \ \
C=\frac{4}{r}\, e^{\phi}\sqrt{N}w',\ \ \ \
\beta=\frac{4}{r}\, e^{\phi}, \ \ \
\gamma=4\sqrt{2N}\, e^{\phi},
$$
\be                                            \label{3.6}
A^{\pm}=2\sqrt{2N}\, e^\phi\phi'\pm (B+1),\ \ \ \
b^\pm=\beta\sqrt{N}\pm\sqrt{2}(B-1).
\ee

The algebraic equations (\ref{3.3}) and (\ref{3.4})
have non-trivial solutions if only the corresponding
determinants vanish:
\be                                             \label{3.7}
A^{+}A^{-}+C^2=0,\ \ \ \ b^{+}b^{-}-w^2\beta^2=0,
\ee
under which conditions the solutions are
\be                                         \label{3.8}
\Psi^1=\frac{A^{-}}{C}\, \Psi^2,\ \ \ \
\Psi^4=\frac{A^{-}}{C}\, \Psi^3,\ \ \ \
\ee
for Eqs. (\ref{3.3}), and
\be                                         \label{3.9}
\Psi^1=\frac{w\beta}{b^{-}}\, \Psi^2,\ \ \ \
\Psi^4=\frac{w\beta}{b^{-}}\, \Psi^3,\ \ \ \
\ee
for Eqs. (\ref{3.4}), respectively. It is clear that these
solutions agree if only
\be                                       \label{3.10}
A^{-}b^{-}=w\beta\, C.
\ee
We thus arrive at the three consistency conditions given by
Eqs. (\ref{3.7}) and (\ref{3.10}), under which the solution
of the algebraic equations (\ref{3.3}) and (\ref{3.4}) 
is expressed by Eqs. (\ref{3.8}) and (\ref{3.9}) 
in terms of two independent functions $\Psi^2$ and
$\Psi^3$. Next, inserting this solution into Eq. (\ref{3.5})
gives an additional consistency condition
\be                                         \label{3.11}
\gamma C\left(\frac{A^{-}}{C}\right)^{\prime}+
2(B+1)A^{-}-A^{-\, 2}-C^2=0,
\ee
and a pair of differential equations for $\Psi^2$ and $\Psi^3$
\be                                         \label{3.12}
\gamma(\Psi^2)^\prime +(A^{-}-B-1)\Psi^2=0,\ \ \ \
\gamma(\Psi^3)^\prime +(A^{-}-B-1)\Psi^3=0.\ \ \ \
\ee
Remarkably, it can be verified that the condition  in Eq. (\ref{3.12})
is a differential consequence of the algebraic conditions (\ref{3.7})
and (\ref{3.10}). The latter therefore provide the full
set of the consistency conditions, under which the solution
of the supersymmetry constraints is given by Eq. (\ref{3.8})
(or Eq. (\ref{3.9})) and Eq. (\ref{3.12}).

Taking into account the definitions in Eq. (\ref{3.6}), the
consistency conditions (\ref{3.7}) and (\ref{3.10}) can be
explicitly expressed as follows:
\begin{equation}                              \label{3.13}
N=1+\frac{r^{2}}{8}e^{-2\phi }
\left(1+2e^{2\phi }\frac{w^{2}-1}{r^2}\right)^2,
\end{equation}
\begin{equation}
r\phi ^{\prime }=\frac{r^{2}}{8N}e^{-2\phi }\left( 1-4e^{4\phi }\,
\frac{(w^{2}-1)^{2}}{r^{4}}\right) ,  \label{3.14}
\end{equation}
\begin{equation}
rw^{\prime }=-2w\frac{r^{2}}{8N}e^{-2\phi }\left( 1+2e^{2\phi }
\frac{w^{2}-1}{r^{2}}\right).  \label{3.15}
\end{equation}
Together with
\begin{equation}
N\sigma ^{2}=e^{2(\phi -\phi _{0})}  \label{3.16}
\end{equation}
these equations provide the full set of the consistency conditions
under which the supersymmetry constraints have non-trivial solutions.
It can be verified that these conditions are compatible
with the field equations (\ref{13}). One can check with the help
of Eqs. (\ref{3.14}) and (\ref{3.15}) that the expression 
for $N$ given by Eq. (\ref{3.13})
is equivalent to that in Eq. (\ref{15}).

The supersymmetry Killing spinors
are given by Eqs. (\ref{3.1})--(\ref{3.1b}) with
\be                                                 \label{3.17}
R^{1}_1=\varepsilon_1 F_1 +\varepsilon_2 F_2 ,\ \
R^{1}_3=\varepsilon_1 F_1 -\varepsilon_2 F_2  ,\ \
R^{3}_2=\varepsilon_1 F_2 +\varepsilon_2 F_1 ,\ \
R^{3}_4=\varepsilon_1 F_2 -\varepsilon_2 F_1  ,\ \
\ee
where
\be                                                   \label{3.18}
F_2 =\exp\left\{-\frac{\phi}{2}
-\int_{0}^r\frac{\sqrt{N-1}}{r\sqrt{N}}dr\right\},\ \ \ \
F_1 =\left.\left.\frac{F_2}{w}\, \right\{e^\phi
\left(\sqrt{N}-\sqrt{N-1}\right)
-\frac{r}{\sqrt{2}}\right\},
\ee
and $\varepsilon_1$, $\varepsilon_2$ are two
complex integration constants.
One can see that there are altogether
four independent Killing spinors.

The same supersymmetry conditions arise in the case where
$g_2=A^{(2)\, a}_{\mu}=0$, whereas 
$A^{(1)\, a}_{\mu}\neq 0$, $g_1=1$. 
Then there are also four independent  Killing spinors.
We therefore conclude that 
the Bogomol'nyi equations (\ref{3.13})--(\ref{3.16})
specify the N=1 supersymmetric BPS states 
in the   N=4 gauged supergravity
with the gauge group SU(2)$\times\left[{\rm U(1)}\right]^3$.

Let us describe briefly what happens 
in the two other cases, where the gauge
symmetry is either Abelian or totally non-Abelian. 
If $g_1=g_2=0$, we make the gauge fields in Eq. (\ref{9}) 
Abelian by setting $w_1=w_2=0$: 
\be                                                \label{3.19}
\T_{(\C)\, a}\, A_{ \mu }^{(\C)\, a}dx^{\mu }=
\T_{(\C)\, 3}\,\cos \theta \,d\varphi, 
\ee
which corresponds to the Dirac monopole type fields. 
The supersymmetry constraints
split then into four independent groups, one group for each 
of the four spinors $\bar{\epsilon}^{\rm I}$. The 
spinors $\bar{\epsilon}^{\rm 1}$ and $\bar{\epsilon}^{\rm 3}$
can be chosen to be independent, 
$\bar{\epsilon}^{\rm 2}$ and $\bar{\epsilon}^{\rm 4}$ being
their Majorana conjugates.  The separation of the 
angular variables is achieved by choosing 
\be                                             \label{3.20}
\bar{\epsilon}^{\, 1}=
\left(
R^{1}_1(r) _{-\frac{1}{2}}Y_{\frac{1}{2}\, m},\ 
R^{1}_2(r) _{-\frac{1}{2}}Y_{\frac{1}{2}\, m},\
R^{1}_3(r) _{\frac{1}{2}}Y_{\frac{1}{2}\, m},\
R^{1}_4(r) _{\frac{1}{2}}Y_{\frac{1}{2}\, m}
\right),\ \ \
\ee
and similarly for $\bar{\epsilon}^{\, 3}$. It turns out then that
if one of the two gauge fields in Eq. (\ref{3.19}) vanishes, 
no matter which, 
the supersymmetry constrains admit two independent solutions
for the for radial amplitudes $R^{1}_q$, and similarly for $R^{3}_q$,
provided that the following consistency conditions hold: 
\be                                    \label{3.21}
N\sigma ^{2}=e^{2(\phi -\phi _{0})},\ \ \ \ \
2Nr^2\phi^{\prime 2}=\frac{e^{2\phi}}{r^2},\ \ \ \
N(1+r\phi')^2=1. 
\ee
In addition, the fact that the azimuthal quantum 
number $m$ in Eq. (\ref{3.20}) assumes two values, $m=\pm1/2$,
doubles the number of solutions, which
finally corresponds to eight supersymmetries. 
The solutions to Eqs. (\ref{3.21}) describe well-known magnetic 
dilaton black holes \cite{G}, the fact that they have N=2
supersymmetry was established in \cite{LK}. 

Finally, in the totally non-Abelian case the supersymmetry
constraints are given by the most general expressions described
above. Similarly to the Abelian case, the minimal value of the 
angular momentum required by the condition (\ref{2.15})
is 1/2. This is due to the presence of the two independent isospins,
which ensures that the total angular momentum is half-integer. 
However, the equations in this case do not
allow for any non-trivial solutions. 

Summarizing, 
the gauged SU(2)$\times$SU(2) N=4 supergravity admits no supersymmetric 
solutions at all  -- in the static, spherically symmetric, purely magnetic sector.  
The ``half-gauged'' SU(2)$\times\left[{\rm U(1)}\right]^3$ model
has solutions with N=1 supersymmetry that will be presented below. 
The non-gauged theory admits solutions with N=2 supersymmetry
described in \cite{G}, \cite{LK}.

\section{Solutions of the Bogomol'nyi equations}
\setcounter{equation}{0}

In order to find the general solution of the Bogomol'nyi equations
(\ref{3.13})--(\ref{3.16}) we start
from the case where $w(r)$ is constant.
The only possibilities are $w(r)=\pm 1$ or $w(r)=0$.

For $w(r)=\pm 1$ the Yang-Mills field is a pure gauge.
Eq. (\ref{3.15}) requires then that $\exp(-2\phi)=0$,
which means that $\phi(r)=\phi_0\rightarrow\infty$, implying
that the metric is flat.

The $w(r)=0$ choice
corresponds to the Dirac monopole gauge field.
The only non-trivial equation Eq. (\ref{3.14}) then reads
\be                                                \label{4.1}
r\phi'=\frac{r^2-2e^{2\phi}}{r^2+2e^{2\phi}},
\ee
whose general solution is  given by
\be                                                \label{4.2}
\phi+\ln\frac{r}{r_0}=\frac{r^2}{4} e^{-2\phi},
\ee
with constant $r_0$.
The corresponding metric turns out to be singular both at the origin
and at infinity.

Suppose now that $w(r)$ is not a constant. Introducing the new variables $
x=w^{2}$ and $R^2=\frac{1}{2}r^{2}e^{-2\phi }$, Eqs.
(\ref{3.13})--(\ref{3.15})
become equivalent to one differential equation
\begin{equation}                                      \label{4.3}
2xR\, (R^2+x-1)\, \frac{dR}{dx}+(x+1)\,R^2+(x-1)^{2}=0.
\end{equation}
If $R(x)$ is known,
the radial dependence of the functions, $x(r)$
and $R(r)$, can be determined from
(\ref{3.14}) or (\ref{3.15}). Eq. (\ref{4.3}) is
solved by the following substitution:
\begin{equation}                                        \label{4.4}
x=\rho^{2}\,e^{\xi (\rho)},\ \ \ \ \ \ \
R^2=-\rho\frac{d\xi (\rho)}{d\rho}-\rho^{2}\,e^{\xi
(\rho)}-1,  \
\end{equation}
where $\xi (\rho)$ is a solution of
\begin{equation}                                       \label{4.5}
\frac{d^{2}\xi (\rho)}{d\rho^{2}}=2\, e^{\xi (\rho)}.
\end{equation}
The most general (up to reparametrizations)
solution of this equation which ensures that $R^2>0$
is $\xi (\rho)=-2\ln\sinh(\rho-\rho_0)$.
This gives us the general solution of  Eqs. (\ref{3.13})--(\ref{3.16}).
The metric is non-singular at the origin if only
$ \rho_{0}=0 $, in which case
\begin{equation}                                      \label{4.6}
R^{2}(\rho)=2\rho\coth \rho-\frac{\rho^{2}}{\sinh ^{2}\rho}-1\, .
\end{equation}
One has $R^{2}(\rho)=\rho^{2}+O(\rho^{4})$ as $\rho\rightarrow 0$,
and $R^{2}(\rho)=2\rho+O(1)$ as $\rho\rightarrow \infty $.
The last step is to obtain $r(s)$ from Eq. (\ref{3.15}), which finally
gives us a family of completely regular solutions
of the Bogomol'nyi equations:
\begin{equation}                                    \label{4.7}
d{ s}^{2}=\left.\left.
2\, e^{2\phi}\,
\right\{
dt^{2}-d\rho^{2}-
\left.\left.R^{2}(\rho)\right(d\vartheta ^{2}+\sin ^{2}\vartheta d\varphi
^{2}\right)\right\} ,
\end{equation}
\begin{equation}                                    \label{4.8}
w=\pm \frac{\rho}{\sinh \rho},\ \ \ \
e^{2\phi }=a^2\, \frac{\sinh \rho}{2\,R(\rho)},
\end{equation}
\begin{figure}
\epsfxsize=10cm
\centerline{\epsffile{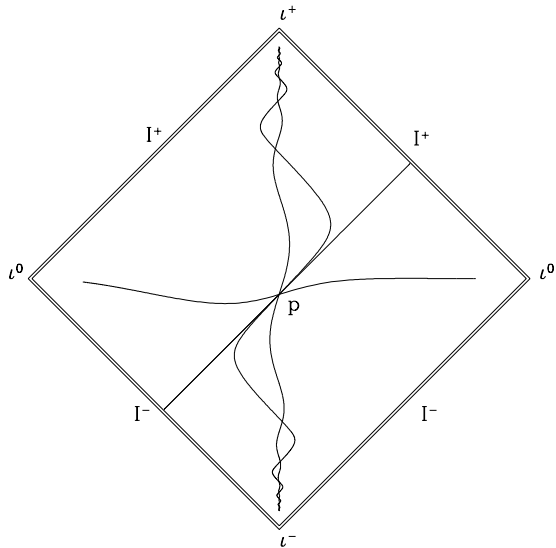}}
\caption{The conformal diagram for the spacetime described by the
line element (5.7).}
\end{figure}

\noindent
where $0\leq \rho<\infty $, and we have chosen in Eq. (\ref{3.16})
$2\phi_0=-\ln 2$. The appearance of the free parameter
$a$ in the solutions reflects the scaling symmetry of
Eqs. (\ref{3.13})--(\ref{3.16}): $r\rightarrow ar$,
$\phi\rightarrow\phi+\ln a$.
The geometry described by the line element (\ref{4.7}) is
everywhere regular,
the coordinates covering the whole space whose
topology is R$^4$. It is instructive to express the solutions
in Schwarzschild coordinates, where the metric functions $N(r)$
and $\sigma(r)$ are given parametrically by
\be                                        \label{4.9}
r=a\sqrt{R(\rho)\sinh\rho},\ \ \
N=\frac{\rho^2}{R^2(\rho)},\ \ \ \
\sigma=\frac{r}{\rho}.
\ee
At the origin, $r\rightarrow 0$, one has
\be                                    \label{4.9:1}
N=1+\frac{r^2}{9a^2}+O(r^4),\ \ \ \ \
N\sigma^2=2e^{2\phi}=a^2+\frac{2r^2}{9}+O(r^4),\ \ \
w=1-\frac{r^2}{6a^2}+O(r^4),\ \ \ \ \
\ee
whereas in the asymptotic region, $r\rightarrow\infty$,
\be                                    \label{4.9:2}
N\propto\ln r,\ \ \ \ \
N\sigma^2=2e^{2\phi}\propto\frac{r^2}{4\ln r},\ \ \ \ \
w\propto\frac{4\ln r}{r^2}.
\ee
The geometry is flat at the
origin, but asymptotically it is not flat.
Specifically, all curvature invariants vanish in the asymptotic
region, however, not fast enough.
For example, the  non-vanishing Weyl tensor invariant
$\Psi_2\propto -1/6r^2$ as $r\rightarrow\infty$.

The global structure of the solutions is well illustrated by the
conformal diagram. Inspecting the $t$-$\rho$ part of the metric,
it is not difficult to see that the conformal
diagram in this case is actually identical to the one for
Minkowski space, even though the geometry is not asymptotically flat
(see Fig.1). The spacetime is therefore geodesically complete and
globally hyperbolic. The latter property is quite remarkable, since
global hyperbolicity is usually lacking for
the known supersymmetry backgrounds in gauged supergravity models.
The geodesics through a spacetime point $p$
are shown in the diagram, each geodesic approaching infinity for
large absolute values of the affine parameter.
Although the global behavior of geodesics  is
similar to that for Minkowski space,
they locally behave differently.
For $\rho<\infty$ the cosmological term $\Lambda(\phi)$ is non-zero
and negative, thus having the focusing effect on timelike geodesics,
which makes them oscillate around the origin.
Unlike the situation in the anti-de Sitter case, each geodesic
has its own period of oscillations,
such that the geodesics from a point $p$
never refocus again.

The shape of the amplitude $w(\rho)$ in Eq. (\ref{4.8})
corresponds to the gauge field of
the regular magnetic monopole with unit magnetic charge. 
In fact, assuming for a moment that 
$\rho$ is the standard radial coordinate, the amplitude
exactly coincides with that for the flat space BPS  solution.
This result is quite surprising, since the model has no
Higgs field, in which case it would be natural to expect the existence of only
neutral solutions \cite{BK}. A manifestation of this is the fact that,
without a Higgs field, the magnetic charge has no gauge invariant
meaning and can only be defined for a certain class of gauges. 
In addition, since all fields in the problem are massless,
it is clear that $w$ cannot in fact exhibit  exactly the same 
 behavior as the one for the flat space BPS monopole amplitude. 
Indeed, passing to the physical radial coordinate $r$, 
the amplitude $w$ for $r\rightarrow\infty$ decays polynomially,
and not exponentially; see Eq. (\ref{4.9:2}).

In conclusion, Eqs. (\ref{4.7}), (\ref{4.8}) describe
globally regular, supersymmetric backgrounds of a new type.
The existence of unbroken supersymmetries suggests
that the configurations should be stable,
and we expect that the stability
proof can be given along the same lines as in \cite{AD}.
Being solutions of N=4 quantum supergravity in four dimensions,
they presumably
receive no quantum corrections. On the other hand,
they can be considered in the framework of the
string theory, and then the issue of
string corrections can be addressed.
In order to study this problem,
we first of all need to lift the solutions to ten dimensions.


\section{Compactification of D=10 supergravity on the group manifold}

\setcounter{equation}{0}

Our aim now is to promote the solutions of the four-dimensional supergravity
model obtained above to the solutions of N=1 supergravity in ten dimensions.
This would make it possible to link the solutions to string theory. It  is a
well-known fact that  ungauged N=4 supergravity in four dimensions can be
obtained via toriodal compactification of ten-dimensional supergravity
\cite{Ali}.
Similarly, the gauged supergravity can be obtained by compactification on
the group manifold. This fact is, however, less known, although one could
have conjectured this by studying the compactification of eleven-dimensional
supergravity on the seven sphere \cite{kaluza}. 
Because this is not covered in the
literature we shall  outline below the compactification procedure in some
detail. We shall restrict ourselves to the purely bosonic sector and
describe the reduction of the action and the fermionic supersymmetry
transformations. The corresponding procedure for the full theory, including
fermion interactions, can be derived  similarly but will not be given here.

\noindent
{\bf 1. The action in D=10.--}
The starting point is the bosonic part of the action of N=1 supergravity in
ten dimensions :
\begin{equation}
S_{10}=\int \left( -\frac{\hat{e}}{4}\,\hat{R}+\frac{\hat{e}}{2}\,\partial
_{M}\hat{\phi}\,\partial ^{M}\hat{\phi}+\frac{\hat{e}}{12}\,e^{-2\hat{\phi}}%
\hat{H}_{MNP}\,\hat{H}^{MNP}\right) \,d^{4}x\,d^{6}z\equiv S_{\hat{G}}+S_{%
\hat{\phi}}+S_{\hat{H}}.  \label{5.1}
\end{equation}
The notation is as follows: the hatted symbols are used for the
10-dimensional quantities. Late capital Latin letters stand for the base
space indices $(M,N,P,\,\ldots )$ and the early letters refer to the tangent
space indices $(A,B,C,\,\ldots )$. For space-time indices taking 4 values,
late and early Greek letters denote base space and tangent space indices,
respectively. Similarly, the internal base space and tangent space indices
are denoted by late and early Latin letters, respectively: 
\begin{equation}
\{M\}=\{\mu =0,\ldots ,3;\,m=1,\ldots ,6\},\ \ \ \{A\}=\{\alpha =0,\ldots
,3;\,{\rm a}=1,\ldots ,6\}.\ \   \label{5.2}
\end{equation}
The general coordinates $\hat{x}^{M}$ consist of spacetime coordinates $x^{{%
\mu }}$ and internal coordinates $z^{m}$. The flat Lorentz metric of the
tangent space is chosen to be $(+,-,\ldots ,-)$ with the internal dimensions
all spacelike.
One has $\hat{e}=\left| \hat{e}_{\ M}^{A}\right| $, the metric
is related to the vielbein by $\hat{{\bf g}}_{MN}=\hat{\eta}_{AB}\hat{e}%
_{\ M}^{A}\hat{e}_{\ N}^{B}=
\eta _{\alpha \beta }\hat{e}_{\ M}^{\alpha }\hat{e}%
_{\ N}^{\beta }-\delta _{{\rm ab}}
\hat{e}_{\ M}^{{\rm a}}\hat{e}_{\ N}^{{\rm b}}\,,
$ and the antisymmetric tensor field strength is 
\begin{equation}
\hat{H}_{MNP}=\partial _{M}\hat{B}_{NP}+\partial _{N}\hat{B}_{PM}+\partial
_{P}\hat{B}_{MN}\,.  \label{5.3}
\end{equation}
The internal space spanned by $z^{m}$ is assumed to form a compact group
space. This means that there are functions
$U_{\ m}^{{\rm a}}\,(z)$ subject to
the condition 
\begin{equation}
\left.\left.
\left( U^{-1}\right) _{{\rm b}}^{\ m}
\left( U^{-1}\right) _{{\rm c}}^{\ n}\right(
\partial _{m}U_{\ n}^{{\rm a}}-\partial _{n}U_{\ m}^{{\rm a}}\right)
=\frac{f_{%
{\rm abc}}}{\sqrt{2}}\, ,  \label{5.4}
\end{equation}
where $f_{{\rm abc}}$ are the group structure constants. The volume of the
space is 
\begin{equation}
\Omega =\int \left| U_{\ m}^{{\rm a}}\right| d^{6}z\, .  \label{5.5}
\end{equation}
In particular, we shall be considering the case where the internal space is
the product manifold SU(2)$\times $SU(2). It is convenient to parametrize
then the 6 internal coordinates by a pair of indices: $\{m\}=\{(\C),i\}$,
where ${s}=1,2$ and $i=1,2,3$; similarly for the tangent space coordinates:$%
\{{\rm a}\}=\{(\C),a\}$, $a=1,2,3$. Each of the two $S^{3}$'s admits
invariant 1-forms $\theta ^{(\C)\,a}=\theta _{\ \ \ i}^{(\C)\,a}dz^{(\C)\,i}$%
\thinspace  :
\begin{equation}
d\theta ^{(\C)\,a}+
\frac{1}{2}\, \epsilon _{abc}\,\theta ^{(\C)\,b}\wedge \theta ^{(\C)%
\,c}=0\,.  \label{5:5:1}
\end{equation}
If we choose
\be                                        \label{5:5:2}
U_{\ m}^{{\rm a}}\equiv U_{\ \ \ i}^{(\C)\,a}=
-\frac{\sqrt{2}}{g_{{s}}}\,\theta
_{\ \ \ i}^{(\C)\,a}\,,
\ee
where $g_{{s}}$ are the two gauge coupling constants, then the structure
constants determined by Eq. (\ref{5.4}) will be
\be                                           \label{5:5:3}
f_{{\rm abc}}\equiv f_{abc}^{(\C)}=g_{{s}}\,\epsilon _{abc}\, .
\ee
Similarly, if one of the gauge coupling constants vanishes, say $g_{2}=0$,
the internal space is SU(2)$\times \left[ {\rm U}(1)\right] ^{3}$. Choosing
in this case $g_{1}=1$,
\be                                         \label{5:5:4}
U_{\ \ \ i}^{(1)\,a}=-\sqrt{2}\,\theta _{\ \ \ i}^{(1)\,a}\,,\ \ \ \
U_{\ \ \ i}^{(2)\,a}=\delta _{i}^{a}\,\ \Rightarrow \ \ f_{abc}^{(1)}=\epsilon
_{abc}\,,\ \ \ f_{abc}^{(2)}=0.  
\ee
{\bf 2. The metric and the dilaton.--}
Let us now return to the general parametrization of the internal space. The
dimensional reduction of the action (\ref{5.1}) starts by choosing the
vielbein and the dilaton in the following form:
\[
\hat{e}_{\ \mu }^{\alpha }\,=e^{-\frac{3}{4}\phi }\,
e_{\ \mu }^{\alpha }\,,\ \ \ \
\ \
\hat{e}_{\ \mu }^{{\rm a}}=\sqrt{2\,}e^{\frac{1}{4}\phi }\,
A_{\mu }^{{\rm a}%
}\,,\ \ \ \ \
\]
\be                                            \label{5.6}
\hat{e}_{\ m}^{\alpha }\,=0\,,\ \ \ \ \
\hat{e}_{\ m}^{{\rm a}}=e^{\frac{1}{4}%
\phi }\,U_{\ m}^{{\rm a}}\,,\ \ \ \ \ \hat{\phi}=-\frac{\phi }{2}\,,
\ee
where all quantities on the right, apart from $U_{\ m}^{{\rm a}}$,
depend only
on $x^{\mu }$. One has $\hat{e}=e^{-3\phi /2}\left| U_{\ m}^{{\rm a}}\right|
\,e$. The dual basis is given by
\[
\hat{e}_{\alpha }^{\ \mu }\,=e^{\frac{3}{4}\phi }\,e_{\alpha }^{\ \mu }\,,
\ \ \
\ \ \ \hat{e}_{{\rm a}}^{\ \mu }=0\,,
\]
\be                                       \label{5.6:1}
\hat{e}_{\alpha }^{\ m}\,=-\sqrt{2}\,
e^{\frac{3}{4}\phi }\,e_{\alpha }^{\ \mu
}\,A_{\mu }^{{\rm a}}\,\left( U^{-1}\right) _{{\rm a}}^{\ m}\,,\
\ \ \ \ \hat{e%
}_{{\rm a}}^{\ m}=
e^{-\frac{1}{4}\phi }\,\left( U^{-1}\right) _{{\rm a}}^{\ m}.
\ee
The metric components are obtained from Eq.(\ref{5.6}) :
\be                                      \label{5.6:2}
\hat{\g}_{\mu\nu}=e^{-\frac{3}{2}\phi}\, \g_{\mu\nu}-
2\, e^{\frac{1}{2}\phi}\, A^{\rm a}_{\mu}A^{\rm a}_{\nu}\, ,\ \ \
\hat{\g}_{\mu m}=\sqrt{2}\, e^{\frac{1}{2}\phi}\,
A^{\rm a}_{\mu}U^{\rm a}_{\ m}\,  ,\ \ \
\hat{\g}_{mn}=- e^{\frac{1}{2}\phi}\, U^{\rm a}_{\ m}U^{\rm a}_{\ n}\, ;
\ee
similarly for $\hat{\g}^{\mu\nu}$.
Using these expressions, the application of the standard formulas \cite{Cho}
gives for the gravitational and dilaton terms in the action (\ref{5.1})
\begin{equation}
S_{\hat{G}}+S_{\hat{\phi}}=\Omega \int e\,\left( -\frac{1}{4}\,R+\frac{1}{2}%
\,\partial _{\mu }\phi \,\partial ^{\mu }\phi -\frac{1}{8}\,e^{2\phi
}\,F_{\mu \nu }^{{\rm a}}F^{{\rm a\mu \nu }}+\frac{1}{32}\,e^{-2\phi }\,f_{%
{\rm abc}}^{2}\right) \,d^{4}x,  \label{5.7}
\end{equation}
where 
\begin{equation}
F_{\mu \nu }^{{\rm a}}=\partial _{\mu }A_{\nu }^{{\rm a}}-\partial _{\nu
}A_{\mu }^{{\rm a}}+f_{{\rm abc}}A_{\mu }^{{\rm b}}A_{\nu }^{{\rm c}}\, .
\label{5.8}
\end{equation}
{\bf 3. The two-form.--}
Now, the important role is played by the antisymmetric tensor field. The
corresponding ansatz is 
\begin{equation}
\hat{B}_{\mu \nu }=B_{\mu \nu }\,,\ \ \ \ \hat{B}_{\mu m}=-\frac{1}{\sqrt{2}}%
\,A_{\mu }^{{\rm a}}\,U_{m}^{{\rm a}}\,,\ \ \ \ \hat{B}_{mn}=\tilde{B}_{mn},
\label{5.9}
\end{equation}
where $B_{\mu \nu }=B_{\mu \nu }(x)$, while $\tilde{B}_{mn}$ depend only on $%
z$. Computation of the field strength according to the rule (\ref{5.3})
gives 
\[
\hat{H}_{\mu \nu \rho }=H_{\mu \nu \rho }\equiv \partial _{\mu }B_{\nu \rho
}+\partial _{\nu }B_{\rho \mu }+\partial _{\rho }B_{\mu \nu }\,,
\]
\[
\hat{H}_{\mu \nu m}=-\frac{1}{\sqrt{2}}\,\left( \partial _{\mu }A_{\nu }^{%
{\rm a}}-\partial _{\nu }A_{\mu }^{{\rm a}}\right) \,U_{\ m}^{{\rm a}}\,,
\]
\[
\hat{H}_{\mu mn}=\frac{1}{2}\,f_{{\rm abc}}\,A_{\mu }^{{\rm a}}\,
U_{\ m}^{{\rm %
b}}\,U_{\ n}^{{\rm c}}\,,
\]
\begin{equation}
\hat{H}_{mnp}=\partial _{m}\tilde{B}_{np}+\partial _{n}\tilde{B}%
_{pm}+\partial _{p}\tilde{B}_{mn}.  \label{5.10}
\end{equation}
We require that 
\begin{equation}
\hat{H}_{mnp}=\frac{1}{2\sqrt{2}}\,f_{{\rm abc}}\,
U_{\ m}^{{\rm a}}\,U_{\ n}^{%
{\rm b}}\,U_{\ p}^{{\rm c}}\,.  \label{5.11}
\end{equation}
This relation should be regarded as a system of equations for $\tilde{B}_{mn}
$. One can see that the solution exists in the cases that we are interested
in. Indeed, if the internal space is $S^{3}\times S^{3}$ Eq. (\ref{5.11})
assures that the 3-form $\hat{H}_{mnp}$ is proportional to the 
volume form on $S^{3}\times S^{3}$. Since this form is closed, the
integrability conditions for the system are locally satisfied. On the other
hand, since the volume form is not exact, the solution exists only
locally. However, the gauge invariance 
\begin{equation}
\hat{B}_{mn}\rightarrow \hat{B}_{mn}+\partial _{m}\Lambda _{n}-\partial
_{n}\Lambda _{m}  \label{5.12}
\end{equation}
allows one to globally extend the local solutions by choosing the
non-trivial transition functions in the overlapping regions. A similar
argument applies when one of the manifolds is $T^{3}$. 

The next step is to compute the vielbein projections of the expressions in (%
\ref{5.10}), (\ref{5.11}). The result is 
\[
\hat{H}_{\alpha \beta \gamma }=e^{\frac{9}{4}\,\phi }\,\left( H_{\alpha
\beta \gamma }-\omega _{\alpha \beta \gamma }\right) ,\ \ \ \ \ \hat{H}%
_{\alpha \beta {\rm a}}=-\frac{1}{\sqrt{2}}\, 
e^{\frac{5}{4}\,\phi }\,F_{\alpha
\beta }^{{\rm a}}\,,
\]
\be                                       \label{5.13}
\hat{H}_{\alpha {\rm ab}}=0\,,\ \ \ \ \ \hat{H}_{{\rm abc}}
=\frac{1}{2\sqrt{2%
}}\, e^{-\frac{3}{4}\,\phi }\,\,f_{{\rm abc}}\,, 
\ee
where $F^{\rm a}_{\alpha\beta}=e_{\alpha}^{\ \mu}
e_{\beta}^{\ \nu}F^{\rm a}_{\mu\nu}$
are the tetrad projections of the gauge field tensor, and
$\omega _{\alpha \beta \gamma }$ are the tetrad projections of the
gauge field Chern-Simons 3-form
\be                                          \label{5.14}
\omega _{\mu \nu \rho }=-6\left( A_{[\mu }^{{\rm a}}
\partial _{\nu }A_{\rho
]}^{{\rm a}}+\frac{1}{3}\,f_{{\rm abc}}\,A_{\mu }^{{\rm a}}\,
A_{\nu }^{{\rm b%
}}\,A_{\rho }^{{\rm c}}\right) . 
\ee
Using Eq. (\ref{5.13}) it is now straightforward to compute the last term in
the action (\ref{5.1}):
\begin{equation}
S_{\hat{F}}=\Omega \int e\,\left( -\frac{1}{8}\,e^{2\phi }\,F_{\mu \nu }^{%
{\rm a}}F^{{\rm a\mu \nu }}-\frac{1}{96}\,e^{-2\phi }\,f_{{\rm abc}}^{\,2}+%
\frac{1}{12}\, e^{4\phi }\, H_{\mu \nu \rho }^{\prime }H^{\prime }{}^{\mu \nu \rho
}\right) \,d^{4}x,  \label{5.15}
\end{equation}
where
\[
H_{\mu \nu \rho }^{\prime }=H_{\mu \nu \rho }-\omega _{\mu \nu \rho }\,.
\]
Now, taking advantage of the identity
\begin{equation}
\varepsilon ^{\sigma \mu \nu \rho }\,\partial _{\sigma }H_{\mu \nu \rho }=0
\end{equation}
it is easy to see that the expression
\begin{equation}
-\Omega \int \left( \frac{1}{6}\,\varepsilon ^{\sigma \mu \nu \rho
}\,\partial _{\sigma }{\bf a}\,H_{\mu \nu \rho }\right) d^{4}x  \label{5.16}
\end{equation}
vanishes up to a surface term; here ${\bf a}$ is a Lagrange multiplier.
Adding this to the action (\ref{5.15}) it is possible to go to a first order
formalism where both  $H_{\mu \nu \rho }$ and  ${\bf a}$ are treated as
independent fields. The equation of motion of ${\bf a}$  implies that $%
H_{\mu \nu \rho }$ is a closed form and can be expressed locally as the curl
of $B_{\mu \nu }$ thus giving the action (\ref{5.15}). Alternatively we can
integrate the field  $H_{\mu \nu \rho }$ from the action as it appears
quadratically. This is equivalent to varying $H_{\mu \nu \rho }$ in the
action with the result
\begin{equation}
H_{\mu \nu \rho }=\omega _{\mu \nu \rho }+e^{-4\phi }\varepsilon _{\sigma
\mu \nu \rho }\,\partial ^{\sigma }{\bf a}\, ,
\end{equation}
and then eliminating  $H_{\mu \nu \rho }$ from the action in favor of ${\bf %
a}$.  Adding Eqs. (\ref{5.7}) and (\ref{5.15}), the result is
\[
S_{10}=\Omega \int e\,\left( -\frac{1}{4}\,R+\frac{1}{2}\,\partial _{\mu
}\phi \,\partial ^{\mu }\phi +\frac{1}{2}\,e^{-4\phi }\,\partial _{\mu }{\bf %
a}\,\partial ^{\mu }{\bf a}-\frac{1}{4}\,e^{2\phi }\,F_{\mu \nu }^{{\rm a}%
}F^{{\rm a\mu \nu }}\right.
\]
\begin{equation}
\left. -\frac{1}{2}\,{\bf a}\,F_{\mu \nu }^{{\rm a}}*\!F^{{\rm a\mu \nu }}+%
\frac{1}{48}\,e^{-2\phi }\,f_{{\rm abc}}^{\,2}\right) \,\,d^{4}x.
\label{5.17}
\end{equation}
Finally, choosing $U_{\ m}^{{\rm a}}$ and $f_{{\rm abc}}$ in accordance with
Eqs. (\ref{5:5:2}) and Eqs. (\ref{5:5:3}), respectively, gives $\left( f_{%
{\rm abc}}\right) ^{2}=6\,(g_{1}^{2}+g_{2}^{2})$, and thus the dimensionally
reduced action (\ref{5.17}) exactly reproduces the bosonic part of the
action of the N=4 supergravity in Eq. (\ref{1}) -- up to an overall factor.
Similarly, the choice (\ref{5:5:4}) leads to the truncated model considered
above.

\noindent
{\bf 4. The fermions.--}
Consider the  supersymmetry transformations for the spinor fields in
ten dimensions (for a purely bosonic background) :
\[
\delta \hat{\psi }_{P}=\hat{D}_{P}\,\hat{\epsilon}+\frac{1}{48}\, e^{-\hat{\phi}%
}\,\left( \hat{\Gamma}_{\ \ \ \ \ \  P}^{MNQ}+9\,\delta _{P}^{M}\,\hat{%
\Gamma}^{NQ}\right) \hat{\epsilon}\,\hat{H}_{MNQ}\,,
\]
\begin{equation}
\delta \hat{\chi }=\frac{i}{\sqrt{2}}\, (\partial _{Q}\hat{\phi})\,
\hat{\Gamma%
}^{Q}\, \hat{\epsilon}+\frac{i}{12\sqrt{2}}\,e^{-\hat{\phi}}%
 \,\hat{\Gamma}^{MNQ}\, \hat{\epsilon}\,\hat{H}_{MNQ}\,.  \label{4a.1}
\end{equation}
Here the D=10 Dirac matrices satisfy $\hat{\Gamma}_{M}\hat{\Gamma}_{N}+\hat{%
\Gamma}_{N}\hat{\Gamma}_{M}=2\,\hat{{\bf g}}_{MN}$, one has $\hat{\Gamma}%
_{M\ldots Q}=\hat{\Gamma}_{[M}\ldots \hat{\Gamma}_{Q]}$. In order to descend
to four dimensions,
we first  notice that for the bosinic background defined by Eqs. (\ref{5.6}%
) and (\ref{5.13}) the vector fields coming from the vielbein and those
from the two-form are identified, while the 36 scalar fields are
truncated. For this to be consistent with supersymmetry, the fermionic fields
which are in the same supermultiplets should also be truncated
simultaneously. In complete analogy with the case of toroidal
compactification one must set  : 
\begin{equation}
\hat{\psi }_{{\rm a}}-\frac{i}{2\sqrt{2}}\,\hat{\Gamma}_{{\rm a}}\,\hat{\chi 
}\equiv 0.  \label{4a.2}
\end{equation}
In order to be consistent with the reduction procedure, 
the variation of the above should remain zero. This
implies that  a Killing spinor $\eta $ exists such that
\be                                                       \label{122}
D_{m}\eta -\frac{g_{{\rm s}}}{4\sqrt{2}}\Gamma _{m}\eta =0\, ,
\ee
where $g_{{\rm s}}=g_{1}$ for $m=1,2,3$, and $g_{{\rm s}}=g_{2%
}$ for $m=4,5,6$. The dependence of the spinors on internal
coordinates  $z$ is factorized through the $\eta $ dependence:
\be                                                       \label{123}
\epsilon \,(x,z)=\epsilon \,(x)\,\eta (z). 
\ee
The next step is to represent the D=10 32-component Majorana-Weyl spinors in
the form 
\begin{equation}
\hat{\psi }_{\alpha }=\left( 
\begin{array}{c}
\psi _{\alpha } \\ 
-i\gamma _{5}\,{\psi }_{\alpha }
\end{array}
\right) ,\ \ \hat{\chi }=\left( 
\begin{array}{c}
{\chi } \\ 
i\gamma _{5}\,\chi 
\end{array}
\right) ,\ \ \hat{\epsilon}=\left( 
\begin{array}{c}
\epsilon  \\ 
-i\gamma _{5}\,\epsilon 
\end{array}
\right) ,\ \   \label{4a.3}
\end{equation}
where $\epsilon \equiv \epsilon ^{{\rm I}}$ with I=1,2,3,4 and $\epsilon ^{%
{\rm I}}$'s are four-component spinors; similarly for $\psi _{\alpha }$ and $%
\chi $. The Dirac matrices are chosen to be 
\begin{equation}
\hat{\Gamma}^{m}=\gamma ^{m}\otimes 1,\ \ \hat{\Gamma}^{1\,a}=\gamma
_{5}\otimes \left( 
\begin{array}{cc}
0 & {\bf T}_{(1)\,a} \\ 
{\bf T}_{(1)\,a} & 0
\end{array}
\right) ,\ \ \hat{\Gamma}^{2\,a}=\gamma _{5}\otimes \left( 
\begin{array}{cc}
-{\bf T}_{(2)\,a} & 0 \\ 
0 & {\bf T}_{(2)\,a}
\end{array}
\right) ,\ \   \label{4a.4}
\end{equation}
where ${\bf T}_{(s)\,a}$ are defined by Eq. (\ref{2.5}). Finally, let us
introduce the following linear combinations: 
\be                                      \label{4a.5}
\psi _{\mu }=e^{-\frac{3}{4}\,\phi }\,\left( e_{\ \mu }^{\alpha }\,\psi
_{\alpha }-\frac{3i}{2\sqrt{2}}\,\gamma _{\mu }\,\chi \right) ,\ \ \ 
\ee
and rescale
\be                                      \label{4a.5a}
\ \chi \rightarrow -2\,e^{-\frac{3}{4}}\,\chi \, .
\ee
The straightforward application of all the above definitions allows one to
verify that the relation between the variations
$\delta \psi _{\mu }$ and $\delta\chi $ of the spinors defined by
Eqs. (\ref{4a.3}), (\ref{4a.5}), (\ref{4a.5a})
and $\epsilon $ in Eq. (\ref{4a.3})
coincides with the D=4 supersymmetry transformation rules in Eq. (\ref{2})
up to the Dirac conjugation. This completes the compactification procedure.


\section{Lifting the solutions to ten dimensions}

\setcounter{equation}{0}

The results of the previous section imply that any solution
of the gauged supergravity model in four dimensions
given in terms of the metric $\g_{\mu\nu}$, 
gauge fields $A^{(\C)\, a}_{\mu}$,
the axion $\A$ and the dilaton $\phi$, can be lifted to ten
dimensions as a solution of the N=1 supergravity.
The ten-dimensional metric, the vielbein
and the dilaton $\hat{\phi}$ are then given by Eqs. (\ref{5.6}) --
(\ref{5.6:2}), where the functions $U^{\rm a}_{\ m}$ are defined
by either Eq. (\ref{5:5:2}) for the SU(2)$\times$SU(2)
 gauge group  or by Eq. (\ref{5:5:4}) when the symmetry is
SU(2)$\times\left[{\rm U}(1)\right]^3$.
If the gauge group is $\left[{\rm U}(1)\right]^6$ one has
$U^{\rm a}_{\ m}=\delta^{\rm a}_{m}$. The vielbein projections
of the three-form are given by Eqs. (\ref{5.13}), from where the
two-form components can be obtained.

Let us now apply these formulas to the family of  solutions obtained in
Section 5. Choosing $A^{(2)\, a}_{\mu}=g_2=0$ and $g_1 =1$,
the lifted solutions can be represented as follows.
The metric and the dilaton are
\be                                                 \label{6.1}
\hat{\g}_{MN}=2 e^{-\hat{\phi}}\, \tilde{\g}_{MN},\ \ \ \ \ \ \
\hat{\phi}=-\frac{\phi(\rho)}{2}\, ,
\ee
where the metric in the string frame, 
$\tilde{\g}_{MN}$, is specified by the line element
\be                                                 \label{6.2}
d\tilde{s}^2 =
dt^{2}-d\rho^{2}-
R^{2}(\rho)\, d\Omega^{2}_2
-\Theta^a\Theta^a
-(dz^4)^2-(dz^5)^2-(dz^6)^2\, .
\ee
Here $d\Omega^{2}_2$ is the standard metric on unit 2-sphere, 
\be                                                \label{6.2:1}
\Theta^a\equiv A^a- \theta^a=
A^{a}_{\mu}\, dx^\mu - \theta^{a}_{\ i}\, dz^i\, ,
\ee
where $\theta^a$ are the Maurer-Cartan forms on $S^3$ parametrized
by $\{z^i\}=\{z^1,z^2,z^3\}$:
\be
d\theta^a+\frac{1}{2}\, \epsilon_{abc}\, \theta^b\wedge\theta^c=0\, .
\ee
If $\T_a$ are the SU(2) group generators,
$[\T_a,\T_b]=i\epsilon_{abc}\T_c\, $, then the gauge field  is given by
\be                                                \label{6.2:2}
A\equiv\T_a A^{a}\equiv
\T_{a}A_{ \mu }^{a}dx^{\mu }=
w(\rho)\,
\{-\T_{2}\,d\theta +\T
_{1}\,\sin \theta \,d\varphi \}+\T_{3}\,\cos \theta \,d\varphi\, .
\ee
The non-vanishing vielbein projections
of the antisymmetric tensor field are
\be
\hat{H}_{\alpha\beta a}=                        \label{6.3}
-\frac{1}{2\sqrt{2}}\, e^{-\frac{3}{4}\phi}\, F^{a}_{\alpha\beta},\ \ \ \
\hat{H}_{abc}=
\frac{1}{2\sqrt{2}}\, e^{-\frac{3}{4}\phi}\, \epsilon_{abc}\, ,
\ee
where $F^{a}_{\alpha\beta}$ are the tertad projections of the
gauge field tensor corresponding to the gauge field (\ref{6.2:2})
for the tetrad $e_\alpha$ specified by the four-dimensional part of the
string metric (\ref{6.2}). These can be read off from
\be                                              \label{6.3:1}
\frac{1}{2}\,\T_a F^{a}_{\alpha\beta}\, e^\alpha\wedge e^\beta=
-\T_2\, \frac{w'}{R}\, e^1\wedge e^2+
\T_1\, \frac{w'}{R}\, e^1\wedge e^3+
\T_3\,\frac{w^2-1}{R^2}\, e^2\wedge e^3\, .
\ee
Finally, for the sake of completeness, we write down 
the functions
$R(\rho)$, $w(\rho)$ and $\phi(\rho)$ in Eqs. (\ref{6.1})--(\ref{6.3:1}) :
\be                                                 \label{6.2:3}
R^{2}=2\, \rho\, \coth \rho-\frac{\rho^{2}}{\sinh ^{2}\rho}-1,\ \ \ \
w=\pm \frac{\rho}{\sinh \rho},\ \ \ \
e^{2(\phi-\phi_0)}= \frac{\sinh \rho}{2\, R(\rho)}\, ,
\ee
where $\phi_0$ is a free parameter. 

One can verify that the lifted solutions given by
Eqs. (\ref{6.1})--(\ref{6.2:3}) indeed fulfill
the equations of motion of ten-dimensional supergravity:
\be                                                      \label{6.4}
\hat{\nabla}_M\hat{\nabla}^M\, \hat{\phi}=
-\frac{1}{6}\, e^{-2\hat{\phi}}\, \hat{H}_{MNP}\hat{H}^{MNP}\, ,
\ee
\be                                                      \label{6.5}
\hat{\nabla}_M\left(e^{-2\hat{\phi}}\, \hat{H}^{MNP}\right)=0\, ,
\ee
\be                                                      \label{6.6}
\hat{R}_{MN}=2\, \partial_M\hat{\phi}\, \partial_N\hat{\phi}
+e^{-2\hat{\phi}}\, \hat{H}_{MPQ}\hat{H}_{N}^{\ \  PQ}
-\frac{1}{12}\, e^{-2\hat{\phi}}\, \hat{\g}_{MN}\,
 \hat{H}_{PQS}\hat{H}^{PQS}\, .
\ee
The direct verification is, however, rather difficult.
Although the dilaton equation can be checked straightforwardly,
already for the antisymmetric tensor field the procedure is much more
involved. The equations then split into three groups depending
on values of the indices $N$ and $P$ in (\ref{6.5}).
Equations of the first group are satisfied by virtue
of the geometrical properties of the invariant forms $\theta^a$,
whereas equations of the second and the third groups
eventually reduce to the Yang-Mills equations in D=4.
Finally, we have had computer check the Einstein equations (\ref{6.6}).

Note that the gauge potential $A$ in Eq. (\ref{6.2:2})
can be arbitrarily gauge transformed, since
any gauge transformation can now be viewed as a diffeomorphism
in ten dimensions.  It is instructive to see how it works
at the linearized level. Consider an infinitesimal gauge
transformation
\be                                               \label{6.2:4}
A\rightarrow A + d\xi + i[\xi,A]\, ,
\ee
where $\xi=\T_a\xi^a(x)$. Consider at the same time
a diffeomorphism
\be                                              \label{6.2:5}
z^i\rightarrow z^i+\theta_{a}^{\ i}(z)\,\xi^a(x)\, ,
\ee
where $\theta_{\ i}^{a}\theta_{b}^{\ i}=\delta^{a}_{b}$, and
the remaining seven coordinates are intact.
This causes a change in the Maurer-Cartan form
$\theta\equiv\T_a\theta^a$:
\be                                               \label{6.2:6}
\theta\rightarrow \theta + d\xi + i[\xi,\theta]\, .
\ee
As a result one has
\be                                                  \label{6.2:7}
\Theta^a\Theta^a=\left.\left.2\,
{\rm tr}\, \right(A-\theta\right)^2\rightarrow
\left.\left.2\, {\rm tr}\,
\right(A-\theta+i[\xi,A-\theta]\right)^2\
=\Theta^a\Theta^a + O\left(\alpha^2\right).
\ee
The D=10 metric therefore remains invariant, and the same
can be shown to be true for the antisymmetric tensor field.
This shows that the effect of gauge transformations can be
compensated by that of the diffeomorphisms.

Finally, let us describe some properties of the solutions in 
Eqs. (\ref{6.1})--(\ref{6.2:3}). 
They preserve 1/4 of the supersymmetries and differ
essentially  from all other known solutions
of leading order string theory \cite{duff} in that the gauge field,
which now appears as off-diagonal components of the metric,
is non-Abelian. For this reason we call the solutions non-Abelian. 
Specifically, the gauge field in the metric combines 
with the non-Abelian isometries of the internal space. 
At first glance, the solutions
exhibit some similarities with  $p$-branes in D=10. 
Here $p=3$ because the expressions do not depend on three
spatial coordinates $z^4$, $z^5$, $z^6$. However, the analogy
is incomplete, since there is no 5-form to couple to the 3-brane. 
In addition, the six-dimensional transverse space is not asymptotically
flat and topologically is R$^3\times S^3$, which spoils the resemblance
with an extended object moving through the 
ten-dimensional spacetime. 
Moreover, we can not introduce the notion of mass of the 
brane per unit 3-volume.

One can regard the solutions as describing interpolating solitons \cite{GT}. 
The reason for this is the observation that for small $\rho$ one
can choose the gauge where the gauge field vanishes in the limit
$\rho\rightarrow0$, and the geometry in string frame
is described by the standard metric on ${\cal M}^7\times S^3$,
where ${\cal M}^7$ is seven-dimensional Minkowski spacetime. 
In the opposite limit, $\rho\rightarrow\infty$, introducing the
radial coordinate 
$\tilde{r}=\sqrt{2\rho}$, the geometry is given by the metric
on ${\cal M}^4\times{\cal V}^6$.
Here ${\cal V}^6$ is a manifold whose metric is 
a ``warped'' product of the standard metric on $S^3$
and that on the three-dimensional paraboloid:
\be                                            \label{6.2:8}
ds^2=\tilde{r}^2\left(d\tilde{r}^2+d\vartheta^2
+\sin^2\vartheta\, d\varphi^2\right)+
\delta_{ab}\, 
(\theta^a-\delta^{a}_3\, \cos\vartheta\, d\varphi)\, 
(\theta^b-\delta^{b}_3\, \cos\vartheta\, d\varphi)\, .
\ee
Note that this does not correspond to any known supergravity vacuum. 

Although we have not studied the issue of $\alpha^\prime$ 
corrections for our solutions, we expect them to get corrected. 
These corrections could probably be balanced by adding the ten-dimensional Yang-Mills field \cite{renata}, however, the definite conclusion can not be
reached without  special analysis. 
This issue is currently under investigation. 
Another interesting problem to analyse is the study of  
dual partners to the solutions found here.

\section{Summary}

In this paper we have studied non-Abelian BPS solutions in  N=4
gauged supergravity and leading order string theory. 
Our main motivation for this was to develop a systematic
procedure for handling non-Abelian gauge fields in the context
of supergravity models, a problem  not well covered in the 
literature. The procedure we have 
employed is  the straightforward component analysis of the equations for  
Killing spinors. 
Although the procedure is rather involved 
(we had to resort to computer calculations) it has given as a set of the 
first integrals (\ref{3.13})--(\ref{3.16})  the field equations 
(\ref{13}) in the static, spherically symmetric, purely magnetic case
with the gauge group SU(2)$\times\left[{\rm U(1)}\right]^3$. 
These first order Bogomol'nyi equations are considerably easier to solve
than the second order field equations, with the solutions
 given by Eqs. (\ref{4.7}), (\ref{4.8}). 

Having obtained the solutions, we show that the N=4 gauged
supergravity
 in four dimensions,
can be obtained via compactification of N=1, D=10 supergravity
on the group manifold. This fact, although quite plausible, 
has not been covered in the literature before. Applying a procedure
inverse to  dimensional reduction, we have lifted the D=4 solutions
to ten dimensions, where they can be regarded as solutions to the 
leading order equations of motion of the string effective action. 

We expect our results to be applicable in the following ways. First, we are
currently investigating the properties of the solutions obtained above
by performing the stability analysis in four dimensions and studying
the issue of string corrections and the duality transformations in D=10. 
Second and more important, we expect that our approach can be applied 
to obtain more general solutions,  also in the context of other 
supergravity models. An interesting example would be N=2 supergravity 
with non-Abelian  matter in four dimensions.


\section*{Acknowledgments}

AHC would like to Nicola Khuri and the Center for Studies in Physics
and Biology at Rockefeller
university for hospitality where a part of this work was done. MSV thanks 
Norbert Straumann for discussions and acknowledges the support of 
the Swiss National Science Foundation and of the Tomalla
Foundation.

\end{document}